\documentstyle[prl,aps,psfig,multicol]{revtex}
\begin{document}

\title{Modulational instability in periodic quadratic nonlinear materials}

\author{J.\ F.\ Corney and O.\ Bang}

\address{Dept. of Informatics and Mathematical Modelling,
Technical University of Denmark \\ DK-2800 Kongens Lyngby, Denmark}

\date{\today}
\maketitle

\begin{abstract}
We investigate the modulational instability of plane waves in quadratic
nonlinear materials with linear and nonlinear quasi-phase-matching gratings.
Exact Floquet calculations, confirmed by numerical simulations, 
show that the periodicity can drastically alter the gain spectrum but never completely removes the instability.  The low-frequency part of the gain spectrum is accurately predicted by an averaged theory and disappears for certain gratings.  The high-frequency part is related to the inherent gain of the homogeneous non-phase-matched material and is a consistent spectral feature.
\end{abstract}

\begin{multicols}{2}
\newlength{\figwidtha}
\newlength{\figwidthb}
\newlength{\figwidthc}
\setlength{\figwidtha}{0.45\linewidth}
\setlength{\figwidthb}{0.95\linewidth}
\setlength{\figwidthc}{0.85\linewidth}

The modulational instability (MI) of waveforms is a fundamental phenomenon
in nonlinear media and is closely associated with the concept of 
self-localized waves, or solitons.  
Like solitons, MI occurs due to an interplay between nonlinear and dispersive or diffractive effects.   
It involves the exponential growth of weak perturbations
through the amplification of sideband frequencies, which causes a destructive 
modulation of the carrier wave.  Many diverse physical systems exhibit MI, e.g.\ fluids\cite{BenFei67}, plasmas\cite{Has75}, nonlinear optics\cite{Ost67}, molecular chains\cite{KivPey92}, and Fermi-resonant interface waves and waveguide arrays\cite{MilBan98}.
In nonlinear optics MI may appear as a transverse instability that 
breaks up a broad optical beam, thereby acting as a precursor for
the formation of stable bright spatial solitons \cite{Fueetal97}.
Conversely, the stable propagation of dark solitons relies on the 
stability of the constant-intensity background and thus requires 
the absence of MI \cite{TriFer95}.

Here we study optical media with a purely quadratic ($\chi^{(2)}$) nonlinearity. 
These materials are of significant technological interest due their strong and fast cascaded 
nonlinearities \cite{chi2review}, which can support stable bright solitons in all dimensions \cite{BerMezRasWyl95PelBurKiv95}.
Moreover, they provide the most general means of studying nonlinear 
processes, because varying the phase mismatch between the fundamental and second-harmonic (SH)
waves changes the nonlinearity from being distinctly quadratic 
to effectively cubic.  
The generality is further enhanced when the properties of the medium are 
modulated with long-period quasi-phase-matching (QPM) gratings. This periodicity not only allows one to tune the effective mismatch, but it also induces effective cubic 
nonlinearities \cite{ClaBanKiv97,CorBan01}, which may be engineered 
to different strengths and signs\cite{BanBalChrTor}.

Except in special cases, MI is unavoidable in $\chi^{(2)}$ materials.  Optical pulses can be modulationally stable only if the dispersion has opposite sign at the fundamental and SH frequencies\cite{BurKiv95}. Transverse inhomogeneities, as in $\chi^{(2)}$ waveguide arrays, may also eliminate MI \cite{MilBan98}.  However, CW beams propagating in pure homogeneous $\chi^{(2)}$ materials are always unstable \cite{TriFer95}.  Naturally, a cubic (Kerr) nonlinearity is always present in $\chi^{(2)}$ materials, and if defocusing and sufficiently strong, it may eliminate MI \cite{AleBurKiv98}. Unfortunately, the cubic nonlinearity in conventional $\chi^{(2)}$ materials is usually focusing, though it may be a factor in the recent observation of apparently stable quadratic optical vortex solitons \cite{TraChietal00}.

We consider transverse MI of CW beams in $\chi^{(2)}$ materials whose linear and nonlinear properties are both modulated and make a detailed study of the instability gain spectra.  Because such spectra can now be measured in the laboratory\cite{FanMalSchSte}, studying their profiles has direct experimental relevance.  Our purpose is to identify regimes with no MI, or with a maximum gain that is small enough to allow spatial dark solitons to propagate stably over experimentally relevant distances.  It is known that, in a system with cubic nonlinearity, periodic dispersion\cite{SmiDor96} (as in dispersion-managed fibers) or nonlinearity\cite{Vla76} can reduce the growth rate of fluctuations.  
We find that, in quadratic materials, the periodicities also drastically alter the MI gain spectrum but they do not entirely remove the instability.

We numerically find exact plane-wave solutions and apply Floquet theory 
to determine their MI gain spectrum.
Numerical simulations, with exact plane-wave solutions seeded with noise
as initial conditions, confirm the Floquet gain spectra.
In the regime of efficient quasi-phase-matching, we find that the MI gain spectrum contains two distinct and well-separated features with fundamentally different physical origins.
The low-frequency part of the gain, which to our knowledge has not been predicted before, is accurately predicted by an averaged theory and disappears for certain grating modulations.  The high-frequency part of the spectrum is related to the 
inherent gain of the non-phase-matched material (i.e., with no gratings) and appears to be unavoidable.  
However, because they are consistently small, the high-frequency peaks can be ignored under a less stringent definition of {\em experimental} stability.  This gives the possibility of stable plane waves as predicted by the averaged theory.

\begin{figure}
  \centerline{\hbox{\psfig{figure=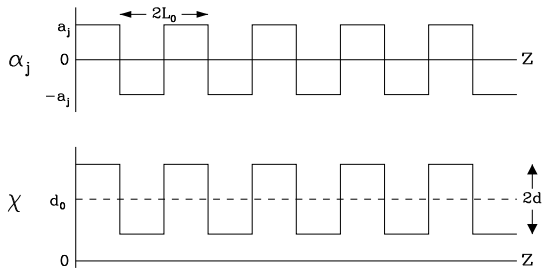,width=\figwidthb}}}
  \vspace*{-0.5cm}
  \caption{Normalized linear and quadratic nonlinear gratings, 
 $\alpha_j(Z)$ and $\chi(Z)$, with domain length $L_0$=$\pi/|k_g|$.} 
  \label{modulation}
\end{figure}

The system under consideration consists of a CW beam (carrier frequency $\omega$) and its SH propagating in a lossless 1D slab waveguide under conditions for 
QPM type-I second-harmonic generation (SHG).
Both the linear and quadratic nonlinear susceptibilities are periodic along 
the $Z$-direction of propagation, as illustrated in Fig.\ \ref{modulation}, but are constant along the transverse $X$-direction.
The modulation of the refractive index is assumed to be weak: $\Delta n_j(Z)/
\bar{n}_j$$\ll$1, where $n_j(Z)=\bar{n}_j+\Delta n_j(Z)$ and $j$ refers 
to the frequency $j\omega$. 
For many materials and etching techniques, such a linear grating appears 
inevitably when the nonlinear susceptibility is altered \cite{FejMagJunBye92}.
We consider gratings for forward QPM only, such that the grating period is 
much longer than the optical period and we can neglect Bragg reflections.
The evolution of the slowly varying beam envelopes is then given by
\cite{MenSchTor94Ban97}
\begin{eqnarray}
  i\partial_Z E_1 + \frac{1}{2} \partial_X^2 E_1 + \alpha_1(Z) E_1 + 
  \chi(Z)E_1^*E_2 {\rm e}^{i\beta Z} & = & 0 ,\nonumber\\
  i\partial_Z E_2 + \frac{1}{4} \partial_X^2 E_2 + 2\alpha_2(Z) E_2 + 
  \chi(Z)E_1^2 {\rm e}^{-i\beta Z} & = & 0, 
  \label{field_eqns}
\end{eqnarray}
where $E_1$=$E_1(X,Z)$ and $E_2$=$E_2(X,Z)$ are the envelope functions of the 
fundamental and SH, respectively.
The coordinates $X$ and $Z$ are in units of the input beam width $X_0$ and 
the diffraction length $L_d$=$k_1 X_0^2$, respectively. 
The parameter $\beta$=$\Delta k L_d$ is proportional to the 
mismatch $\Delta k$=$k_2-2k_1$, where $k_j$=$j\omega\bar{n}_j/c$ is the 
average wave number. Thus $\beta$ is positive for normal dispersion and 
negative for anomalous dispersion.
The normalized refractive-index grating is $\alpha_j(Z)$=$L_d\omega\Delta 
n_j(Z)/c$, and the normalized nonlinear grating is $\chi(Z)$=$L_d\omega 
d_{\rm eff}(Z)/(\bar{n}_1 c)$, where $d_{\rm eff}$=$\chi^{(2)}/2$ is given 
in MKS units.

To consider a plane wave with longitudinal wave number offset $\Lambda$ 
and transverse wave number $\Omega$, we transform to the new variables 
$w(x,z)$=$E_1(X,Z)\exp{(i\Omega X-i\Lambda Z)}$, $v(x,z)=E_2(X,Z)\exp
(2i\Omega X-2i\Lambda Z+i\tilde\beta Z)$, $x$=$\sqrt{\eta}(X+\Omega Z)$,
and $z$=$\eta Z$, where $\eta=|\Lambda +\frac{1}{2}\Omega^2|$ for 
$\Lambda\neq -\frac{1}{2} \Omega^2$ and $\eta$=1 otherwise.  We have also introduced a residual effective mismatch $\tilde\beta$=$\beta-k_g$.
This gives the equations
\begin{eqnarray}
  i\partial_zw + \frac{1}{2}\partial_x^2 w - rw + \alpha_1'(z)w 
        + \chi'(z)w^*v {\rm e}^{i\kappa z} &=& 0 \nonumber \\
  i\partial_zv + \frac{1}{4}\partial_x^2v - \sigma v + 2\alpha_2'(z)v
        + \chi'(z) w^2 {\rm e}^{-i\kappa z} &=& 0 ,
  \label{planewave2}
\end{eqnarray}
where $r = {\rm sgn}(\Lambda +\frac{1}{2}\Omega^2)$, 
$\sigma = 2r-\tilde\beta/\eta$, and $\kappa = k_g/\eta$.
Note that $r$=0 when $\Lambda$=$-\frac{1}{2}\Omega^2$. 
We Fourier-expand the rescaled gratings, $\alpha_j'(z)$=$\alpha_j(Z)/\eta$ and $\chi'(z)=
\chi(Z)/\eta$:
\begin{equation}
  \label{FourGrating_F}
  \alpha_j' = a_j' \sum_n g_n {\rm e}^{in\kappa z} , \;
  \chi'     = d_0' + d' \sum_n g_n {\rm e}^{in\kappa z},
\end{equation}
where, for the square waveform that we consider (Fig.\ \ref{modulation}), $g_n$=$2s/(i\pi n)$ for $n$ odd and $g_n$=0 for $n$ even.
The sign $s$=${\rm sgn}(\kappa)$ is included to ensure that the 
expansions give the same grating form for both signs of $\kappa$.  
The gratings will drive a periodic variation in $z$ of spatial frequency $\kappa$ in the fields of system (\ref{planewave2}).  We explicitly include this fast variation by expanding the fields in Fourier series also:
\begin{equation}
  \label{FourField_F}
  w = \sum_n w_n(x,z) {\rm e}^{in\kappa z} , \;
  v = \sum_n v_n(x,z) {\rm e}^{in\kappa z}.
\end{equation}
For plane-wave solutions $(w,v)$=$(w_s,v_s)$, all the coefficients $w_n$ and 
$v_n$ are constants determined by
\begin{eqnarray}
  -(n\kappa+r) w_n + a_1'\sum_lg_{n-l}w_l + \sum_{l,p}D_{n+l-p-1}w_l^*v_p
  &=&0, \nonumber \\
  -(n\kappa+\sigma) v_n + 2a_2'\sum_lg_{n-l}v_l + \sum_{l,p}D_{n-l-p+1}w_lw_p
  &=&0, \nonumber
  \label{F}
\end{eqnarray}
where $D_n$=$d'g_n+d_0'\delta_n$, $\delta_n$ being Kronecker's delta.
We find the solution numerically using a standard relaxation technique 
based on Newton's method.

We determine the linear stability of the plane-wave solutions by 
using exact Floquet theory
\cite{Fueetal97}.  For perturbations of transverse frequency offset $\nu$:
\begin{eqnarray}
  w(x,z) &=& w_s(z) + \delta w_1(z){\rm e}^{-i\nu x} + \delta w_2^*(z)
  {\rm e}^{i\nu x}, \nonumber \\
  v(x,z) &=& v_s(z) + \delta v_1(z){\rm e}^{-i\nu x} + \delta v_2^*(z)
  {\rm e}^{i\nu x}, 
\end{eqnarray} 
the linear growth is governed by the linearized equation $\partial_z{\bf P} = M(z){\bf P}$, where ${\bf P}(z)$ and the matrix $M(z)$ are
\begin{eqnarray}
  {\bf P} = \left( \begin{array}{c}
  \delta w_1\\ \delta w_2\\ \delta v_1\\ \delta v_2 
  \end{array} \right),\;  M = i\left(\begin{array}{cccc}
  a & b & c^* & 0 \\ -b^* & -a & 0 & -c \\ 
  2c & 0 & d & 0 \\ 0 & -2c^* & 0 & -d \end{array}\right),
\end{eqnarray}
and where we have defined the components $a = -\nu^2/2 - r + \alpha_1'(z)$, $b = v_s(z)\chi(z){\rm e}^{i\kappa z}$, $c = w_s(z)\chi(z)e^{-i\kappa z}$ and $d = -\nu^2/4 - \sigma + 2\alpha_2'(z)$.

Because of the periodicity in $\alpha_j'$, $\chi'$, $w_s$ and $v_s$, the matrix $M$ is also periodic with period $z_p=2\pi/|\kappa|$.  
To determine the stability, we need only find the eigenvalues 
$\lambda_i$ of the mapping $\delta T$ of the solution over one period: 
${\bf P}(z+z_p)=\delta T{\bf P}(z)$.  
We calculate $\delta T$ numerically by integrating the linearized equation \cite{Fueetal97} with a fourth-order Runge-Kutta method.  
If all the $\lambda_i$ lie on the unit circle for every 
$\nu$, then the solutions $w_s$ and $v_s$ are stable; otherwise  
there is a gain whose profile is $g$=$\max\{\Re[\ln(\lambda_i)/z_p]\}$.  We note that the actual growth rate at short distances may differ from $g$, depending on the overlap of the initial perturbation with the unstable eigenvector\cite{Ros96}.  However, for larger distances ($z >1/g$), as we consider here, the growth rate will approach the limiting maximum value $g$.

\begin{figure}
  \centerline{\hbox{
  \psfig{figure=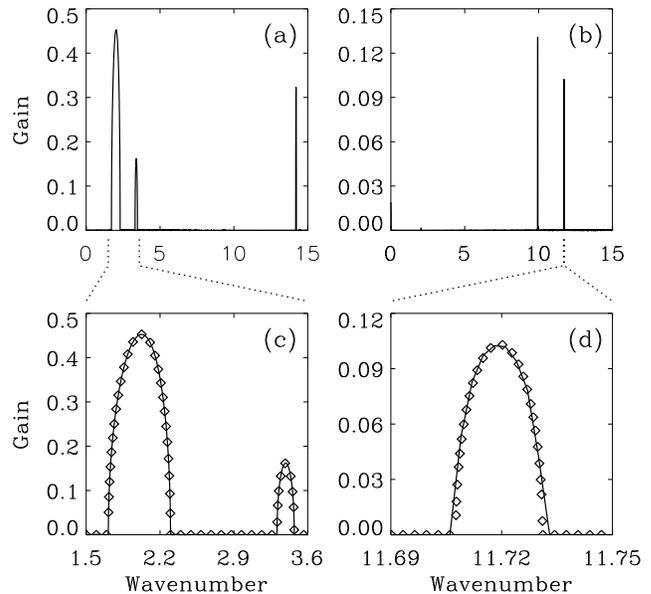,width=\figwidthc}}}
\vspace*{0.5cm}  
\caption{Floquet gain spectra for (a) the LiNbO$_3$ and (b) the GaAs/AlAs gratings, for $\sigma$=$-2$. The diamonds give the averaged-field results in (c) and the equivalent non-phase-matched homogeneous results in (d).}
  \label{gain_1}
\end{figure}

As a first example, we calculate the Floquet gain spectrum for a conventional domain-reversal grating in LiNbO$_3$ ($a_1'$=$a_2'$=$d_0'$=0) for $r$=$-1$, $\kappa$= 100 and exact phase matching ($\sigma$=$-2$).  As Fig.\ \ref{gain_1}(a) shows, the spectral features fall into two well-separated regions: two low-frequency bands around $\nu$=2.0 and $\nu$=3.4, and a narrow high-frequency band at $\nu$=14.2.  For comparison Fig.~\ref{gain_1}(b) plots the gain for a GaAs/AlAs superstructure with a nonlinear grating etched though quantum-well disordering\cite{HelHutetal00} ($d_0'/d'$=4.6) and a linear grating chosen to be $(a_1'-a_2')/\kappa$=$-1/9.2$.  High-frequency gain bands are again present, but the low-frequency bands are absent.  Further results show that such narrow, high-$\nu$ bands are a general feature 
of the gain spectrum for QPM gratings.  They are related to the inherent MI in the non-phase-matched, gratingless $\chi^{(2)}$ material:  Each gain peak that appears in the gratingless spectrum usually also appears in the Floquet spectrum, typically with the same height and spectral location.  An exception is the gain from the symmetric grating case ($d_0'$=0), as in LiNbO$_3$, where the original `homogeneous' peak is suppressed.  Often, as in Fig.\ \ref{gain_1}(b), this peak is split into two or more closely spaced components by the grating.

\begin{figure}
  \centerline{\hbox{\psfig{figure=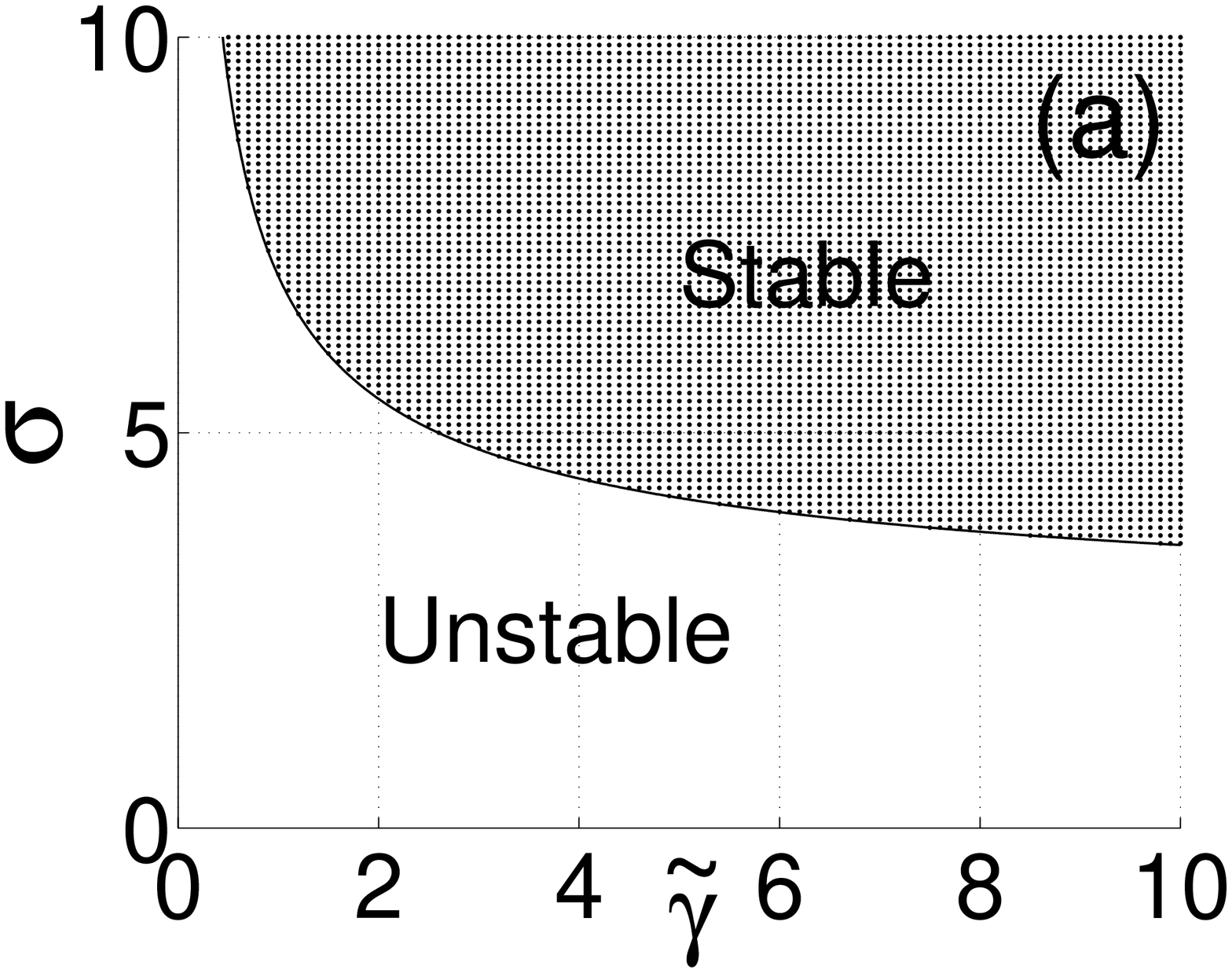,width=\figwidtha}
  \psfig{figure=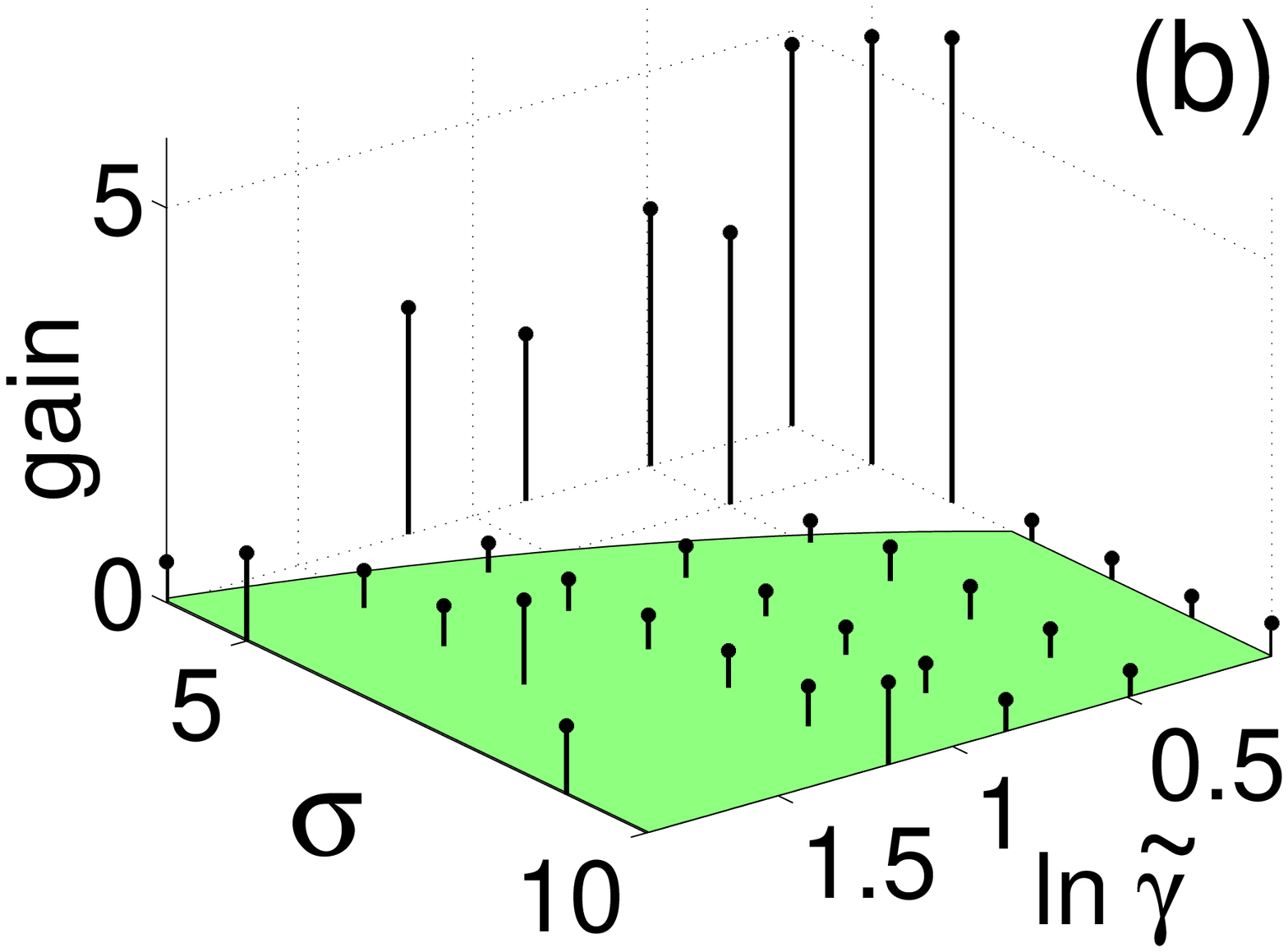,width=\figwidtha}}}
  \caption{(a) Region of MI predicted by the 
  averaged-field theory with $r$=$-1$ and $\rho$$\ne$0. (b) Maximum gain predicted by Floquet theory for 5$\le$$\kappa$$\le$30, with $d_0'/d'$=5/3 and $(a_1'-a_2')/\kappa$=0.158.  The region of zero gain predicted by averaged-field theory is shaded.} 
  \label{stability}\label{pl_stab}
\end{figure}

The low-frequency features in the gain spectrum are accurately predicted by approximate 
averaged-field equations for the DC components $(\bar w,\bar v)=(w_0,v_0)$.  
From Eqs.~(\ref{planewave2}) these equations are found to be\cite{CorBan01,ClaBanKiv97}
\begin{eqnarray}
   &&i\partial_z \bar w + \frac{1}{2}\partial_x^2\bar w - r\bar w 
     + \rho \bar w^*\bar v + \gamma(|\bar v|^2-|\bar w|^2)\bar w = 0, 
     \nonumber\\
   &&i\partial_z \bar v + \frac{1}{4}\partial_x^2 \bar v - \sigma\bar v 
     + \rho^* \bar w^2 + 2\gamma|\bar w|^2\bar v = 0,
  \label{average_eqns}
\end{eqnarray}
where $\rho=2is(d'+2d_0'(a_1'-a_2')/\kappa)/\pi$ and $\gamma=
(d_0'^2+d'^2(1-8/\pi^2))/\kappa$.  The averaged equations are valid for small residual mismatch ($\beta' \ll \beta$) and small grating period ($\kappa \gg 1$).  They show that
the linear grating may decrease the strength of the effective quadratic 
nonlinearity, possibly even to zero \cite{CorBan01,FejMagJunBye92}.
More importantly, the nonlinear grating induces cubic nonlinearities.  These can be defocusing
($\gamma>0$) \cite{CorBan01} and may thus stabilize plane waves \cite{AleBurKiv98}.
We find that Eqs.~(\ref{average_eqns}) indeed have modulationally stable 
plane-wave solutions in a large regime for $r=-1$, as shown in 
Fig.\ \ref{stability}(a), where $\tilde\gamma=\gamma/|\rho|^2$ gives the 
relative strength of the cubic nonlinearity.  The figure also shows that, since the mismatch $\sigma$ cannot be large (otherwise the averaged equations are invalid), the effective quadratic nonlinearity $|\rho|$ must be reduced (i.e., $\tilde\gamma$ increased) to get a stable solution.

The case shown in Fig.\ \ref{gain_1}(a) corresponds to having  $\tilde\gamma=44.5$, $\sigma = -2$, which lies outside the region of stability predicted by the averaged equations.  The gain profile, shown in Fig.\ \ref{gain_1}(c), matches exactly the low-frequency part of the Floquet spectrum.  The case in Fig.\ \ref{gain_1}(b) corresponds to $\rho = 0$, in which case the averaged equations reduce to two coupled Schr\"odinger equations.  The plane-wave solution 
$w_0$=$\sqrt{-r/\gamma}$, $v_0$=0 is stable for $\gamma$$>$0 
($r$=$-1$), in agreement with the absence of low-frequency bands in the exact Floquet spectrum.

\begin{figure}
  \centerline{\hbox{\psfig{figure=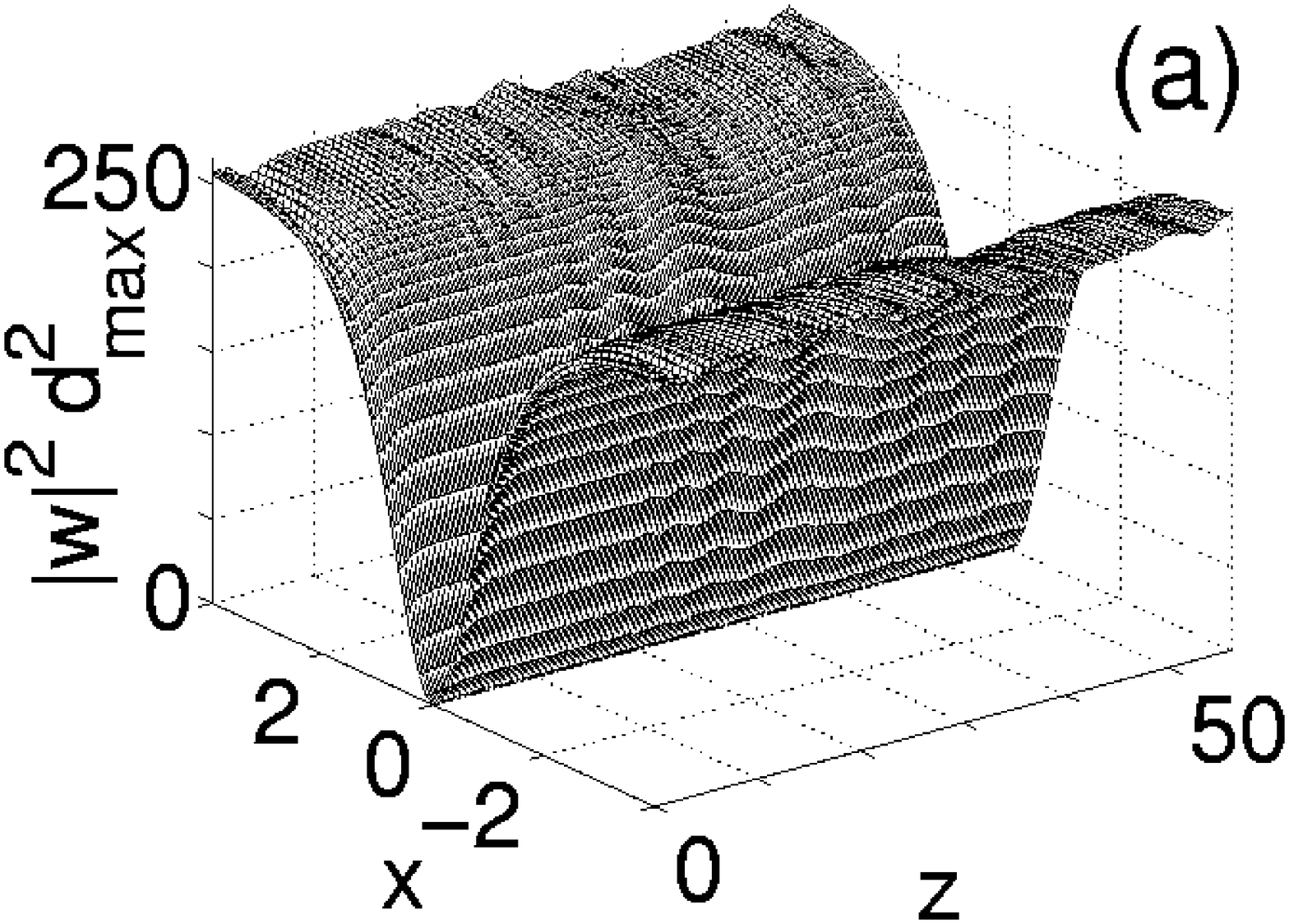,width=\figwidtha}
  \psfig{figure=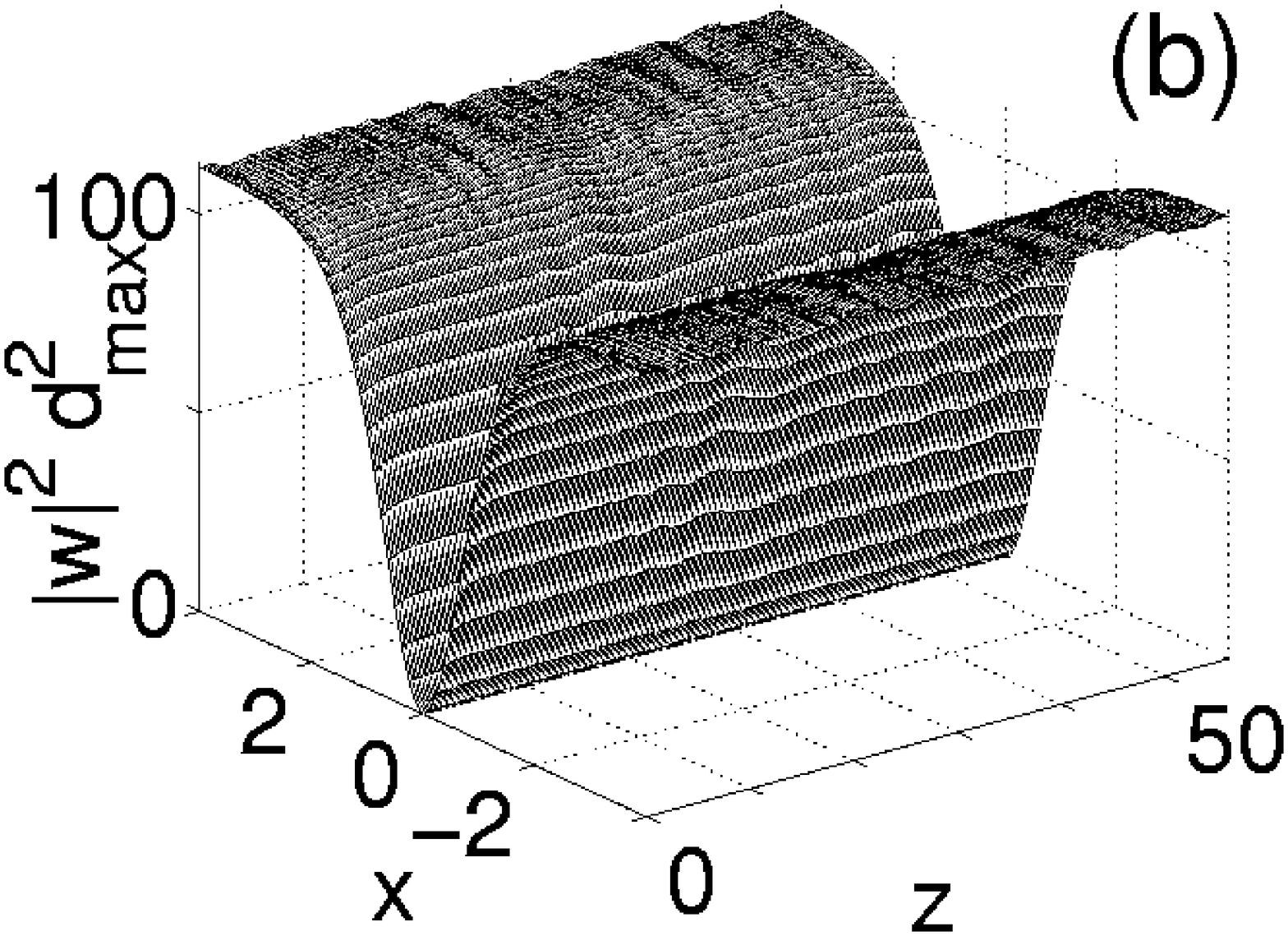,width=\figwidtha}}}
  \caption{Propagation of dark solitons with (a) a modulated nonlinearity ($d_0'/d' = 4.6$, $\kappa = 112$) and (b) no modulation ($d' = 0$).  In both cases $\beta = 100$ and $L_d \sim 200 \mu m$.}
  \label{dark_sol}
\end{figure}

Figure \ref{pl_stab}(b) shows the maximum gain calculated from Floquet theory for a given modulation depth.  Where the averaged equations predict MI, the Floquet gain is large, in agreement with the averaged results.  However where the averaged equations predict stability, Floquet theory finds a residual gain, due to the high-frequency gain bands.  
However, because the residual gain is often small and narrow, it may 
be neglected under a reasonable definition of {\em experimental stability}, 
which allows for a small growth over say 10 diffractions lengths (typical 
length scale in experiments). 
Neglecting the high-frequency gain bands causes the results of Floquet and 
average-field calculations to coincide and whole regions in parameter space of ``experimentally 
stable'' plane-waves to appear.  These regions are even larger than shown in Fig.~\ref{stability} and also appear for $r$=$+1$ and $r$=0.  We demonstrate this experimental stability by simulating the propagation of dark solitons with Eq.\ (\ref{field_eqns}), shown in Fig.\ \ref{dark_sol}(a).  The parameters correspond to a nonlinear grating induced in GaAs/AlAs through quantum-well disordering\cite{HelHutetal00}, as in Fig.\ \ref{gain_1}(b) but with no linear modulation.  Stable propagation of the dark soliton can be seen for over 60 diffraction lengths.  Now because of the small gain in highly non-phase-matched materials, such stable propagation can also be achieved without the nonlinear grating (Fig.\ \ref{dark_sol}(b)). But the SHG efficiency is then very low, as evidenced by the relative SH intensity being a factor of 10 less than with the grating.

\begin{figure}
  \centerline{\hbox{
  \psfig{figure=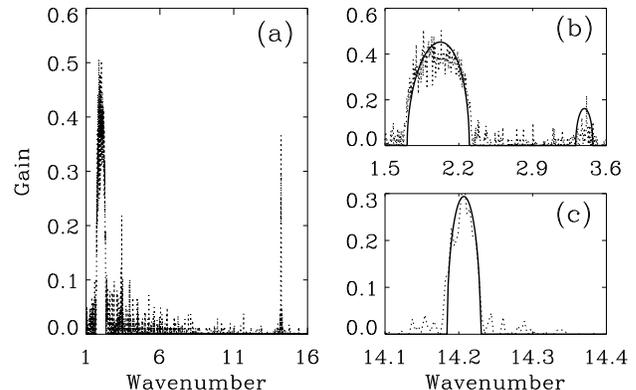,width=\figwidthc}}}
  \vspace*{0.6cm}
  \caption{(a) Gain profile calculated from propagative simulations, for the same parameters as in Fig.\ \ref{gain_1}(a). Comparisons with the Floquet theory (solid curve) are given in (b) and (c). }
  \label{gain_curve3}
 \label{prop3}
\end{figure}

The Floquet spectral results are confirmed by numerical simulations 
of Eqs.\ (\ref{field_eqns}).  
We launch plane waves seeded with Gaussian white noise to excite all spatial frequencies.
The gain spectrum is then calculated from the Fourier-transformed 
fundamental $\tilde w$ as $\tilde g(\nu)=(\ln|\tilde w(z_2,\nu)| 
- \ln|\tilde w(z_1,\nu)|)/(z_2-z_1)$ \cite{HeArrSteDru99}.
As shown in Fig.\ \ref{prop3}, the simulations agree with the Floquet curves, in both the low-$\nu$ (Fig.\ \ref{prop3}(b)) and high-$\nu$ (Fig.\ \ref{prop3}(c)) regions.

In summary, we have used an exact Floquet technique to find the MI gain spectrum in $\chi^{(2)}$ materials with general QPM gratings.  Due to the periodicity, novel high- and low-frequency gain bands appear in the spectrum. A simple approximate averaged-field theory correctly predicts the low-frequency gain bands.  The high-frequency peaks that are consistently present in the spectrum are typically small and may be neglected under a realistic definition of experimental stability.  The predictions of the averaged and the exact Floquet analyses then agree, and stable plane waves and dark solitons become possible, even under conditions close to phase matching.

MI is yet to be investigated in $\chi^{(2)}$ materials with both linear 
and nonlinear QPM gratings.
However, it is now possible to directly probe the gain spectrum in
experiments \cite{FanMalSchSte}, and thus the interesting spectral 
profiles predicted here may be verified in the laboratory.

This work is supported by the Danish Technical Research Council under grant no. 5600-00-0355.

\end{multicols}

\end{document}